\documentclass[aps,prc,twocolumn,superscriptaddress]{revtex4}
\usepackage{graphicx,amsmath,amssymb,bm}

\newcommand{\be}{\begin{equation}}
\newcommand{\ee}{\end{equation}}

\newcommand{\fmi}{\, \text{fm}^{-1}}
\newcommand{\mev}{\, \text{MeV}}
\newcommand{\vlk}{V_{{\rm low}\,k}}
\newcommand{\la}{\Lambda}

\begin{document}

\title{Modern topics in theoretical nuclear physics}
\author{B.K. Jennings}
\email[E-mail:~]{jennings@triumf.ca}
\affiliation{TRIUMF, 4004 Wesbrook Mall, Vancouver, BC, Canada, V6T 2A3} 
\author{A. Schwenk}
\email[E-mail:~]{schwenk@triumf.ca}
\affiliation{TRIUMF, 4004 Wesbrook Mall, Vancouver, BC, Canada, V6T 2A3} 
\affiliation{Nuclear Theory Center, Indiana University, Bloomington, IN 47408}

\begin{abstract}
\centerline{prepared for the Nuclear Physics Town Hall Meeting at
TRIUMF, Sept. 9-10, 2005}\vspace*{4mm}

Over the past five years there have been profound advances in nuclear
physics based on effective field theory and the renormalization group. 
In this brief, we summarize these advances and discuss how they impact our 
understanding of nuclear systems and experiments that seek to unravel 
their unknowns. We discuss future opportunities and focus on modern
topics in low-energy nuclear physics, with special attention to
the strong connections to many-body atomic and condensed matter physics, 
as well as to astrophysics. This makes it an exciting era for nuclear 
physics.
\end{abstract}

\maketitle

\section{Introduction}

Nuclear physics is undergoing a renaissance. This results
from a confluence of novel interests, experimental opportunities and 
significant developments in theoretical nuclear physics. 

First, there is a need for precise nuclear input to astrophysical 
calculations. Much that happens in the sky has a nuclear physics 
underpinning, from how the sun shines to what happens when stars 
explode in a supernova. The formation of the elements, nucleosynthesis, 
requires the knowledge of nuclear masses, lifetimes and 
reactions at low energies. 
Astrophysical interests tied with the quest for an understanding of nuclear 
systems at the limits of stability have motivated the construction of 
radioactive beam facilities worldwide: ISOLDE at CERN, SPIRAL at GANIL,
NSCL at MSU, ISAC at TRIUMF, RIBF at RIKEN (under construction), FAIR 
at GSI (construction begin 2008), RIA (planned with highest 
construction priority) and EURISOL (proposed). With 
these experimental facilities, it is now possible to vary the 
composition of known nuclei to extreme neutron to proton ratios 
and investigate their phenomena. Nuclear physics also uses
many new experimental techniques, such as ion or atom traps for 
high-precision studies. 
In addition, nuclei play an important role in constraining 
the standard model of particle physics, for example the $V_{ud}$ 
parameter in the CKM matrix and the Dirac or Majorana nature of neutrinos.

From a condensed matter perspective, studies of rare isotopes may be regarded 
as ``femtoscience'' research, where one starts with known nuclei and 
studies their dependence on ``doping'' by neutrons or protons. As in
condensed matter physics, some key aspects of novel nuclear systems are 
superfluidity/superconductivity and the phenomena of frustrated 
systems, for nuclei due to the competition of the long-range electromagnetic 
and short-range strong interactions. 

It is readily
anticipated that the frontiers investigated with radioactive beam 
facilities will lead to industrial and medical applications, similar 
to the techniques that stem from prior university-based accelerator 
facilities. Today, one of three hospitalized patients undergoes 
a nuclear medicine procedure.

Astrophysics questions and exciting nuclear experiments
are spurring a new generation of nuclear theorists to 
develop systematic approaches to the nuclear many-body problem. In particular, 
a novel understanding of nuclear interactions 
based on effective field theory and 
the renormalization group plays a crucial role in the theoretical advances.

In this brief, we identify three main aspects of current research in nuclear 
physics. The first is as an extremely rich and complex nuclear many-body 
problem that spans 18 orders of magnitude from nucleons to neutron stars 
and is ultimately based on QCD. It is therefore important to develop a 
reliable and coherent description of nuclear phenomena over this range. 
Since the many-body methods used for nuclei and neutron stars are general,
theoretical progress in nuclear physics can have direct impact for instance 
on the theory of quantum dots and cold atoms. The development of many-body
methods reaches across science:
The coupled cluster technique was first proposed in the 
context of nuclear physics, further developed with many extensions in 
quantum chemistry, and now has returned to nuclear physics for use in 
ab-initio calculations of the properties of medium-mass nuclei. 
Similarly, renormalization group methods that were proposed for 
electronic systems, such as the Hubbard model, have recently been 
applied to nucleonic matter and superfluidity in neutron stars.

Second, nuclear physics is an essential part of astrophysics. There would 
have been no solar neutrino problem, if the nuclear physics and 
hydrodynamics of the sun had not been sufficiently well understood 
to make accurate predictions of the expected neutrino flux. Many other 
astrophysical processes intimately involve nuclear physics. The equation 
of state and the neutrino response of nuclear matter are crucial inputs to 
simulations of supernova explosions and neutron star cooling. Nuclear 
masses, lifetimes and reactions drive nucleosynthesis. Nuclear energy and 
gravity are the prime movers in the universe.

The third aspect of nuclear physics is as a necessary part of experiments
needed for particle physics or other subfields of physics. For example, the
parameter $V_{ud}$ in the CKM matrix can be determined from nuclear $\beta$ 
decay. Here the nucleus can be considered as part of the target holder
of the quark undergoing $\beta$ decay, and clearly a successful experiment 
requires that the relevant nuclear physics be well understood. 

The second and third 
aspects can obviously only take place in the context of the first. 
The role of the 
nucleus in the cosmos cannot be understood without first understanding 
the nucleus in the laboratory.

In this brief, we first discuss how effective field theory and the
renormalization group place nuclear interactions in the hierarchy of theories
that range from atomic and 
condensed matter to high-energy particle physics. These
ideas lead to a change in our understanding of the nuclear 
many-body problem and to the development of novel and refined methods, 
which put nuclear many-body theory in the main stream of general many-body 
physics. The advances also have important consequences for
experiments. Throughout this presentation, we focus on modern topics 
in low-energy nuclear physics, with special attention to the various strong 
connections to atomic, condensed matter and astrophysics.

\section{The Big Picture}

The reductionist goal is to reduce everything to one all-encompassing 
theory. At the Planck scale $M_{\text{P}} \sim 10^{19} \, \text{GeV}$, 
string theory holds that promise. The verification of string theory
is complicated however, since it demands either extremely high-energy 
probes or reliable calculations of very small contributions to low-energy
processes. Between the Plank scale and 
the electro-weak breaking scale, $M_{\text{ew}} \sim 10^{2} \, \text{GeV}$,
little is known. In this energy range there are a number of possible effective 
field theories, such as grand unified theories, supersymmetric theories
or left-right symmetric models.

At the electro-weak breaking scale $M_{\text{ew}}$, the standard model of 
strong and electro-weak interactions is well explored and extraordinarily 
successful. (The dynamic mechanism of electro-weak symmetry breaking
and the nature of the Higgs boson are still open problems.) However, 
it is certainly not the ``theory of everything'', since it does not describe 
gravity. Therefore the question arises: How can a theory be successful, when 
the underlying theory to which it is an approximation is unknown? This is 
where the idea of renormalization comes in. The physics at distances too 
short to be probed is not resolved and can be replaced by simpler 
interactions whose couplings encode all short-distance
contributions. This leads to an effective field theory with a finite number 
of low-energy constants. The masses of the elementary particles are such 
constants. They can either be determined experimentally, or by matching 
amplitudes calculated in the effective field theory and the underlying 
theory. While details of the underlying theory are not needed to make 
predictions at low energies, the higher energy theory may
unify the phenomena of the standard model and reduce the number of its
parameters.

At the next energy scale, we have QCD as the theory of strong interactions. 
QCD can be divided into three regimes. At high energies, $M_\text{pQCD} 
\gtrsim$ few~GeV, QCD becomes asymptotically free and interactions 
of quarks and gluons are perturbative. At intermediate energies, $M_\text{had} 
\sim 1 \, \text{GeV}$, the physics is that of strongly-interacting hadrons,
with a diverse spectrum of mesons and baryons and their resonances. Nuclear
physics governs the interactions of the lightest baryons and the lightest
mesons at energies $M_\text{nuc} \sim 100 \, \text{MeV}$. The nuclear 
physics regime naturally carries a key imprint of QCD, since the pions, 
the lightest mesons, are the Goldstone bosons of chiral symmetry breaking.
Each of the three regimes have their own relevant degrees of freedom and
thus their own effective theory. Effective field theory allows us to separate 
nuclear physics from the more complicated problem of hadronic physics, 
while maintaining a clear and direct connection to QCD.
At lower energies, we have atomic and condensed matter physics, where
nuclear and standard model parameters (nuclear masses, charges, charge radii)
are low-energy constants. The resulting picture of physics thus consists of 
layer upon layer of effective field theories, and all interactions are 
effective interactions in this sense.

The problem at every energy scale decouples from the physics at 
higher energies. The high-energy information is encoded in the low-energy 
coefficients of the effective field theory. This decoupling is very useful, 
even when the effective theory at the higher energies is known, as with
QCD and nuclear physics. This is because the effective field theory 
focuses on the physics of the problem at hand, and provides a method to 
separate out the relevant information from complicated details that are 
not needed. For example, Fermi's theory of $\beta$ decay works well without 
explicitly including the $W$ and $Z$ bosons.

The idea of effective field theory and decoupling run counter to the
reductionist's ideal. The physics at every energy scale can be studied 
largely independent of what happens at other energies. It is not 
necessary to understand interactions at higher energies to make progress
at a given scale. An extreme example is critical phenomena, where the 
critical exponents depend only on very general properties, the symmetries
and the dimensionality, of the bare Lagrangian. All the remaining 
information is integrated out, and the results depend only on the fixed 
points of the renormalization group. It is the renormalization group and 
not the bare couplings that determine the critical exponents. 

As discussed in the next section, 
a similar discovery was recently made in nuclear 
physics: Microscopic approaches to nuclear systems traditionally start
from various models for the inter-nucleon interaction, which differ
substantially at short distances and lead to model dependences for nuclei
and neutron stars. Using the renormalization group, it was shown that all 
these microscopic nuclear forces evolve to a universal nucleon-nucleon 
interaction for nucleons with momenta $k \lesssim 4 M_\text{nuc}$, when 
the model-dependent high-momentum parts are integrated out. This
unifies nuclear structure and nuclear astrophysics calculations. Nuclear
interactions can be thought of as one in the 
sequence of effective interactions that stretch from string theory at 
highest energies through the standard model of particle physics to the 
effective interactions of atomic and condensed matter physics.  

\section{Nuclear Interactions based on EFT and RG}

The goal of nuclear theorists is to provide a coherent and systematic 
description of all low-energy nuclear phenomena, those occurring on
earth and in stars. The theoretical framework can be used to predict nuclear 
processes that are currently not accessible experimentally, and to 
understand the microphysics of observed phenomena. Therefore, nuclei, 
neutron stars and supernovae serve as laboratories to test effective 
interactions at nuclear energies. A rigorous test requires that the 
truncations in the nuclear Lagrangian are well understood and that 
the microscopic many-body calculations are reliable.
In this section, we review the advances in our understanding of nuclear 
interactions based on Effective Field Theory (EFT) and the Renormalization 
Group (RG). We will discuss recent progress in many-body calculations 
in the following section.

The connection of nuclear interactions to QCD is through EFT. When nuclear
systems are probed at low energies, the relevant degrees of freedom are
nucleons and pions, the lightest confined baryons and mesons respectively.
Nuclear EFT includes explicitly nucleon and pion fields and allows all
possible interactions that are consistent with the symmetries of QCD, 
most importantly chiral symmetry. In order to be predictive and systematic, 
an organization (``power counting'') must be present to permit a finite 
truncation of possible terms in the Lagrangian. For nuclear interactions, 
the power counting originally proposed by Weinberg~\cite{Weinberg} is in 
powers of the typical momenta in nuclear systems $Q$ over the EFT breakdown 
scale $\Lambda_\chi$. An estimate for
the low-momentum scale is $Q \sim m_\pi$, where $m_\pi \approx 140 \mev$ 
is the pion mass, and the EFT breakdown scale is $\Lambda_\chi \sim 700 
\mev$, where heavier mesons (like the $\rho$ meson with $m_\rho \approx
770 \mev$) and nucleon resonances are resolved and have to be included
explicitly. The pions are the Goldstone bosons of chiral symmetry breaking, 
and consequently their interactions become weaker at low energies. This
is a key QCD property, which is clearly present in its EFT. The low-energy 
EFT does not include explicitly heavier baryons or quarks and gluons. All
their effects are present in the low-energy couplings of the simpler
short-range interactions. For further details we refer the reader 
to several excellent reviews on EFT applied to nuclear 
forces~\cite{EFTreview1,EFTreview2} and also to the general introduction to EFT
by Lepage~\cite{Lepage}.

Nuclear interactions based on EFT have now been developed to
Next-to-Next-to-Next-to Leading Order, N$^3$LO or up to $(Q/\Lambda_\chi)^4$, 
in the low-momentum expansion~\cite{N3LO,N3LOEGM}. To this order, one has
$24$ low-energy constants, which have been fit to the world set of 
$\sim 4500$ neutron-proton and proton-proton scattering data. A
substantial number of these data have been measured at TRIUMF.
The reproduction of two-body observables is very good
with $\chi^2/\text{datum} \approx 1$. We emphasize that the fit
to experiment automatically takes into account all high-energy
effects on low-energy observables. Alternatively, since nuclear EFT 
is connected to QCD, the low-energy constants can be determined
using lattice QCD or other nonperturbative methods. The continuous
advances in lattice QCD, for example full simulations with light
quarks, make this a long-term vision for nuclear forces. For
exciting progress on nuclear physics from QCD see~\cite{Savage}.

The physics
of nuclear EFT is extremely rich. The pionic interactions lead to 
significant non-central parts in nuclear forces. These are analogous 
to magnetic dipole-dipole interactions and depend on the orientation 
of the nucleon spins relative to their position. In many-body systems,
such interactions can lead to remarkable phenomena, for instance the
differences observed in the A and B phases of liquid $^3$He recognized
with the 2003 Physics Nobel Prize~\cite{Leggett}. The spin-independent 
part of nuclear interactions is similar to molecular van-der-Waals 
potentials, and thus nuclear systems combine the phenomena of spin 
systems and atomic systems.

There are several key advantages of nuclear EFT. In nuclear physics, 
many-body forces are inevitable. (Note that QCD has a three-quark 
interaction through the non-Abelian $ggg$ vertex.)
It is established beyond doubt that for all realistic 
nucleon-nucleon (NN) interactions, a significant three-body force is
required to describe light nuclei~\cite{GFMC1,GFMC2,Nogga,NCSM}. 
In nuclear EFT, the power counting predicts that the first three-nucleon 
(3N) interaction is present at N$^2$LO; their effects are thus suppressed 
by $(Q/\Lambda_\chi)^3$ compared to NN 
interactions~\cite{chiral3NF1,chiral3NF2}. A similar hierarchy 
can be derived for many-body forces using nuclear EFT. The EFT many-body 
forces are consistent with the NN interaction, for example the long-range 
part of the 3N and 4N forces are parameter-free; their couplings also 
enter the NN interaction and are constrained from $\pi N$ or $\pi \pi$
scattering respectively. Moreover, the N$^3$LO 3N force only has two
low-energy constants. 

In addition to a controlled and viable expansion of 
many-body forces, the nuclear EFT approach can address questions like: 
How would nuclear shell structure or nucleosynthesis change, if the up 
and down quark masses would be different? A first investigation of this 
sort was the quark-mass dependence of the binding energy of the 
deuteron~\cite{qmass1,qmass2}, which is
the lightest nucleus and important to nuclear burning.
Moreover, neutrino-nucleon~\cite{nuN} or parity-violating
interactions~\cite{Pviolating} can be consistently incorporated in
applications of nuclear EFT to studies of fundamental symmetries.

\begin{figure*}[t!]
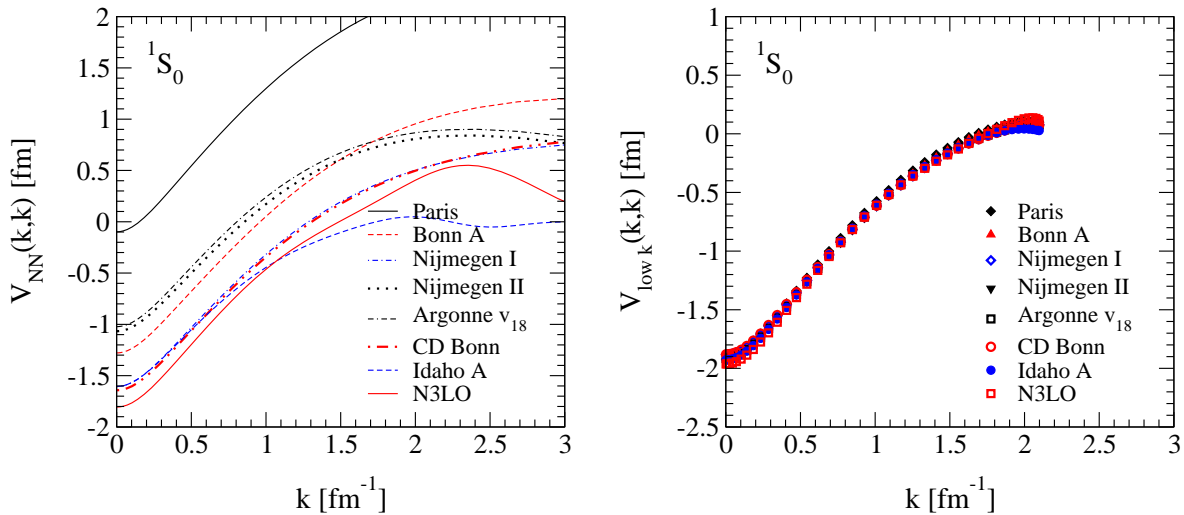

\begin{center}
\includegraphics[scale=0.46,clip=]{vnn_1s0_before.eps}
\hspace*{4mm}
\includegraphics[scale=0.46,clip=]{vlowk_1s0_after.eps}
\end{center}
\vspace*{-4mm}
\caption{\label{Vlowk} Left figure: Different nuclear interaction models 
$V_\text{NN}(k,k)$ versus incoming/outgoing relative momentum $k$. 
Right figure: The resulting low-momentum interactions $\vlk(k,k)$, 
derived by integrating out the high-momentum ($p > 2.1 \, \text{fm}^{-1}$)
modes using the RG. The interactions are shown for the $^1$S$_0$ partial 
wave, but the same universal behavior is found in all partial 
waves and for lower cutoffs~\cite{Vlowk1,Vlowk2}. The momentum units are 
$1 \, \text{fm}^{-1} \approx 200 \, \text{MeV}$.}
\end{figure*}

The second advancement to our understanding of nuclear 
interactions comes directly from the RG. The idea behind renormalization 
is that the effects of the high-momentum modes on the low-momentum theory 
can be simulated by a set of simpler interactions. In the regime of the 
EFT expansion, the simpler interactions are contact interactions 
and their derivatives. Using the RG, it is possible to integrate out
modes with momenta larger than a cutoff $\Lambda$ starting from NN 
interactions. The resulting effective interaction, defined for momenta below
$\la$, encodes all high-energy effects in the low-energy couplings.
While nuclear physics may seem
unfamiliar with the renormalization terminology, the basic ideas of the 
RG are present in the seminal work of Bloch and 
Horowitz~\cite{BH} and in subsequent nuclear many-body methods.

To illustrate the power of the RG in nuclear physics, we review the
application to nuclear interactions~\cite{Vlowk1,Vlowk2}. RG invariance
requires that the low-momentum theory reproduces the same low-momentum
scattering amplitude $T$ as a given large cutoff nuclear interaction with 
high-momentum modes. Therefore, the low-momentum interaction $\vlk$ satisfies
\begin{multline}
T({\bf k'},{\bf k},E_{\bf k}) = \vlk^\la({\bf k'},{\bf k}) \\[1mm]
+ \int^{\la} \frac{d^3 {\bf p}}{(2\pi)^3} \:
\vlk^\la({\bf k'},{\bf p}) \: G^{(2)}({\bf p},E_{\bf k}) \: 
T({\bf p},{\bf k},E_{\bf k}) \,,
\end{multline}
where ${\bf k'}$ and ${\bf k}$ are the incoming and outgoing momenta, the 
scattering energy is $E_{\bf k}$, and $G^{(2)}({\bf p},E_{\bf k})$ denotes 
the two-particle propagator. The cutoff in Eq.~(1) defines the resolution
scale of the effective theory, since details at distances $r \lesssim 1/\la$
are not resolved. The cutoff independence of the scattering amplitude, 
$d T({\bf k'},{\bf k},E_{\bf k}) / d\la = 0$, leads to a RG equation that 
determines how $\vlk$ evolves as the cutoff is lowered and high-momentum 
modes are integrated out. 

In the left part of Fig.~\ref{Vlowk}, we show the N$^3$LO EFT, as well as 
various models for the NN interaction. These potential models all include 
the same long-range one-pion exchange interaction and are fit to scattering
data for momenta $k \lesssim 2.1 \fmi$ (roughly up to the threshold for pion
production), but they differ substantially in their treatment of the
high-momentum physics. When these model dependences are removed using 
the RG, one obtains a universal low-momentum interaction $\vlk$ for all
cutoffs $\la \lesssim 2.1 \fmi$. This is shown in the right part of 
Fig.~\ref{Vlowk}. The RG thus unifies microscopic nuclear forces and
removes these model dependences from many-body observables.

For EFT interactions, the RG can be used to change the resolution scale in
nuclear forces, analogous to evolving the renormalization scale in
parton distribution functions in QCD. The RG evolution of EFT interactions 
to lower cutoffs generates all higher-order short-range interactions 
necessary to maintain RG invariance, and therefore it has been argued 
that $\vlk$ effectively parameterizes a higher-order nuclear EFT 
interaction~\cite{Vlowk3NF}. In summary, instead of a number of models,
we now have inter-nucleon interactions (from EFT and RG) that depend 
on the resolution scale. This is
similar to the running coupling in QCD and to improved lattice actions, 
where the resolution is $\la = \pi/a$, $a$ being the lattice spacing.

For cutoffs below the breakdown scale, the scaling of higher-order
two-body or many-body forces is dominated by the cutoff, because
$Q/\la>Q/\la_\chi$. Many-body observables that do not 
include many-body forces will be cutoff dependent and the cutoff 
variation gives an estimate for the neglected physics.
Varying the cutoff in nuclear interactions is thus a powerful tool to 
assess the truncation errors and the completeness of the calculation. 
Such theoretical 
uncertainties are clearly important for predictions of nuclear 
systems, for terra incognita at ISAC or in stars. They also provide valuable
guidance, where future experimental constraints are most needed.

The RG casts very useful insights into why the different potential 
models have worked as well as they have and shows their limitations.
One criticism of the traditional potential models is that their short-range 
parts should be constructed from explicit quark degrees of freedom. 
They should not be simple interactions, often taken to be local or
approximately local, based on heavy-meson exchanges, dispersion 
theory or in some cases phenomenology.
The EFT and RG understanding, however, shows that it is entirely 
legitimate to replace physics that is not resolved by something 
simpler~\cite{Lepage}. The higher energy physics, including quark 
effects, is taken into account by fitting these simpler interactions 
to experiment. The problem with the potential models is not that 
the short-distance details are incorrect, but more so that they lack 
a systematic scheme to construct consistent many-body 
interactions and effective operators. In addition, the potential models 
have strong high-momentum modes and large cutoffs, and one could be 
misled into thinking a higher cutoff implies that the physics is more 
valid. While the RG approach shows that the different potential models
lead to the same low-momentum theory, it also demonstrates that their 
high-momentum behavior is not constrained by data. Theoretical studies 
that are sensitive to these short-range parts are at best incomplete.

An important result is that 3N interactions become weaker for lower 
cutoffs~\cite{Vlowk3NF}. This shows that nuclear interactions with
large cutoffs are counterproductive, because 3N forces in this case
must cancel loop contributions from high-momentum intermediate states 
that are incorrectly represented. 3N interactions corresponding to 
$\vlk$ are perturbative in light nuclei~\cite{Vlowk3NF} and 
therefore tractable. As
a result, we will be able to perform the first calculations with 3N
forces for intermediate-mass nuclei~\cite{VlowkCC}. Fundamental progress 
in nuclear physics requires a better understanding of 3N interactions.
These lead to particular density and isospin dependences, and experimental 
information from nuclei therefore provides significant constraints. The ISAC
facility can play a unique role in this endeavor.

A weakening of 
the 3N force is also observed empirically in the Unitary Correlation 
Operator Method~\cite{UCOM}, where an explicit unitary basis 
transformation is developed that removes the short-range correlations. 
The resulting effective interaction is similar to $\vlk$, and as in
the RG approach, this implies that short-range correlations are not
strict observables. In addition, large-basis diagonalizations within the
No-Core Shell Model show a similar behavior. As the basis includes more 
high-energy states of the NN interaction, one finds a deterioration
from the experimental binding energies~\cite{NCSM1}.

Effective interactions have a long history in nuclear physics. In 
the Bloch-Feshbach-Horowitz projection operator 
formalism~\cite{BH,Feshbach} (the same framework is used to understand 
and predict Feshbach resonances in cold atoms), the effective interaction 
carries a characteristic energy dependence. For many-body systems, a
correct treatment of this energy dependence is complicated, with promising
progress for light nuclei~\cite{HL}.
The RG approach of Refs.~\cite{Vlowk1,Vlowk2} and equivalent unitary 
transformations~\cite{LS} lead to energy-independent interactions.
Thus the concept of effective interactions is more general than sometimes
realized~\cite{BKJ1}.

In addition to the interplay between NN and 3N interactions, there are
effective operators that are also resolution scale dependent and evolve with 
cutoff. To see why effective operators are necessary, consider for example
electromagnetic interactions. Photons can couple to quarks inside nucleons 
and to heavier mesons/baryons that have been integrated out. Electromagnetic 
operators must include these effects. Nuclear EFT provides a systematic 
scheme to incorporate all contributions with simpler interactions. We stress 
that effective operators and many-body forces cannot be meaningfully 
discussed outside the context of a given two-body interaction.

We conclude this section by discussing nuclear interactions at very low
energies, where even the pion is not resolved. The corresponding EFT is
constructed from contact interactions and their derivatives. Pionless
EFT has been extremely successful for strongly-interacting systems with 
large scattering lengths~\cite{pionless}. 
These systems exhibit universal properties at
low densities, independent of the atomic or nuclear details, because there
are no lengths scales associated with the interaction in this regime.
For example, for two spin states with equal populations, the equation
of state of cold gases of $^6$Li or $^{40}$K atoms in the vicinity of
Feshbach resonances is identical to the equation of state of low-density 
neutron matter in the crust of neutron stars. Applications of the pionless
EFT developed in nuclear physics range from cold atoms~\cite{EricHans}
to nuclear reactions~\cite{haloEFT} and neutron stars~\cite{dEFT}. There
are many nuclear reactions with large scattering lengths and thus progress
in their theoretical description can be directly applied to describe cold
atom gases with similar resonant interactions. 

Finally, the EFT and RG
again provide useful insights why simple nuclear potential models 
developed for nuclear reactions work well. For example, the recent
calculation of the $^7$Be(p,$\gamma$)$^8$B reaction with simple, pionless
interactions is rather successful~\cite{Be7pg}. In the new understanding,
these simpler interactions are nuclear interactions at a lower resolution
scale, which can be thought of as integrating out the pions in nuclear
interactions. The EFT and RG thus provide a way to built on the 
empirical successes of these calculations, in a way that other modeling
does not.

\section{The Nuclear Many-Body Problem based on EFT and RG}

The traditional problem in nuclear physics is that nuclear interactions
are strong and model dependent (see e.g., the discussion in~\cite{Vlowknm}). 
Both aspects can be traced 
to the same sources. First, different models for the strongly repulsive 
core lead to different behavior at high momenta or high virtual energies.
Second, nuclear interactions have tensor forces that scatter strongly to 
high momenta.  The optimal way to deal
with the high-momentum modes is to convert the problem first to a low-momentum
theory more appropriate to the resolution at hand. This realization has lead to
a revolution in nuclear physics. A common feature in
modern nuclear many-body developments is that they first integrate out 
(and thus suppress) high-momentum virtual 
contributions~\cite{Vlowk2,HL,NCSM2,CC,UCOM}.
In the process, low-momentum effective interactions are generated. This is
achieved either directly by the RG~\cite{Vlowk1,Vlowk2} or through
other effective interaction methods~\cite{HL,NCSM2,CC,UCOM} (for comparative
details see~\cite{BKJ2}). This separation of the long-range
physics from the short-range details builds on the separation of scales in the
hadron spectrum of QCD.

The treatment of high-momentum modes has been a continual difficulty in 
nuclear physics.  The
success of Dirac phenomenology~\cite{SW} has been traced to the fact that it
suppresses high loop momenta~\cite{Thies,Cooper}. In pion-nucleus scattering
there is a similar improvement when high-momentum loop contributions are
suppressed~\cite{piN}. As discussed in the previous section, nuclear
interactions with lower cutoffs or those that suppress high-momentum modes have
weaker 3N forces. It is intriguing that these effects correlate with a
weaker iterated pion exchange, and investigations in this direction may
lead to valuable insights.

A significant advantage of low-momentum interactions is that they can be 
directly applied to nuclear many-body systems with model-independent results 
and without uncontrolled resummations~\cite{Vlowknm}. 
For systems with $A < 100$ particles, 
the prime approaches are the nuclear shell model~\cite{SM} and the coupled 
cluster method~\cite{CC}. Presently, exact shell-model diagonalizations are 
possible for all semi-magic nuclei, and for $A<70$ nuclei in $0 \hbar \omega$
space. The coupled cluster method is the method of choice for systems with 
up to $100$ electrons in quantum chemistry. 
First applications of low-momentum interactions 
to the nuclear shell model are very promising~\cite{VlowkSM}. These studies
can provide a basis for future calculations of the exotic structure
of nuclei investigated at ISAC. As discussed, 3N interactions become 
weaker for low-momentum cutoffs and are thus tractable in larger systems.
As a result, we will be able to include microscopic 3N interactions 
beyond the lightest nuclei. The first coupled cluster calculations with 
this endeavor are under way~\cite{VlowkCC}.
 
For bulk properties of nuclear matter, low-momentum 
interactions offer the possibility of a perturbative and therefore systematic 
approach~\cite{Vlowknm}. In contrast to the traditional perception, the 
role of the 3N interaction is essential for saturation. These results
demonstrate that the power counting must change for nuclear 
matter~\cite{Vlowknm}, and the cutoff scaling of different contributions can 
provide guidance for the development of a systematic EFT. In addition, the 
nuclear matter results imply that exchange correlations are tractable, 
and this motives a program to derive the nuclear density functional from 
microscopic interactions.
Density Functional Theory (DFT) is the method of choice to study ground 
state properties of $A > 100$ particle systems~\cite{DFT1,DFT2,DFT3,DFT4}.
The density functional is universal and the 
microscopic foundations of DFT are well-understood. In a path 
integral approach, the density functional is the effective action for the 
density. We refer the reader to~\cite{DFTfound} for a short
discussion of the microscopic foundations and DFT opportunities in the 
context of nuclear physics. Here, we only remark that the nuclear density
functional comes with novel and unique features due to superfluidity and 
the self-bound nature of nuclei. These are largely unexplored for electronic
systems.

In addition, there are promising non-perturbative RG methods for nuclear
many-body systems~\cite{RGnm}, based on the RG approach to interacting 
Fermi systems proposed by Shankar~\cite{Shankar}. For condensed matter
systems, this approach has been applied to study the phase diagram 
of the two-dimensional Hubbard model~\cite{HM1,HM2,HM3} and electronic 
excitations in atoms~\cite{elex}. In nuclear physics, the RG method was
first used to investigate superfluidity in neutron stars~\cite{RGnm}, 
since pairing properties are a pivotal input to simulations of neutron star 
cooling. A systematic extension of the RG method to finite systems will
lead to non-perturbative shell-model interactions and effective operators 
for heavy nuclei.

The long-term vision of nuclear many-body theory is a microscopic and 
predictive understanding of nucleonic matter under extreme compositions, 
temperatures and densities, with theoretical error estimates. There are
many common themes with atomic and condensed matter systems: How does 
the structure of matter change with its composition? What is the many-body 
physics of complex and collective phenomena in nuclear systems? How are 
shape transitions in nuclei and neutron stars related to frustrated
systems and the phase diagram of asymmetric matter? What are the 
connections to mesoscopic systems and chaos? In addition, the EFT 
connection to QCD makes it possible to address
how nuclear many-body phenomena depend on QCD parameters, for example 
the quark masses.

Reliable many-body approaches are especially important for the
nuclear physics input to stars, supernovae and the formation of the
elements. The nuclear equation of state is key to supernova and neutron
stars, where nucleonic matter range from a complex gas at extremely low 
densities to a liquid at nuclear densities. Below saturation density,
nucleonic matter comes in various shapes known as ``nuclear pasta''.
Experiments with low-energy heavy-ion collisions and connections to
frustrated systems can provide insights on the cluster
physics of the complex gas and the nuclear pasta. Recent progress comes
from molecular dynamics simulations of the pasta phases~\cite{pasta1,pasta2} 
and a model-independent description of low-density nuclear matter based 
on the  virial expansion~\cite{vEOS}. The resulting virial equation of 
state constrains the physics of the neutrinosphere in supernovae, and 
differs from all models used in supernova simulations. In addition,
neutrino interactions with nucleonic matter are key to supernova 
explosions and the cooling of neutron stars. The many-body advances
make it possible derive reliable neutrino-matter
interactions consistent with the nuclear equation of state and including
the effects due to nuclear clusters. 
Moreover, nuclear superfluidity strongly modifies the cooling of neutron 
stars, since pairing of neutrons or protons suppresses neutrino emission
via $\beta$ decay and neutrino bremsstrahlung. It is exciting that there 
is a similar observation in nuclei: For the halo nucleus $^{11}$Li, the 
$\beta$ decay occurs in the $^9$Li core, since the pairing energy 
suppresses the decay of the two halo neutrons~\cite{li11}.

Since the nucleon-nucleon scattering lengths are large, the nuclear
many-body problem has an exciting overlap with the universal physics
of cold atoms with resonant interactions. The physics of low-density
neutron matter is part of the BEC-BCS crossover physics studied with 
trapped Fermi gases~\cite{Thomas,Salomon,Jin,Grimm,Ketterle}. 
Likewise, many nuclear reactions have large 
scattering lengths, and a model-independent description of nuclear 
reactions can therefore be used to study other resonant interactions.
Another exciting overlap exists with asymmetric~\cite{asymmetric}
or rotating Fermi gases and asymmetric nuclear matter, rotating nuclei 
or neutron stars.

\section{Conclusions}

Nuclear physics covers the low-energy regime of QCD and explores the 
interactions of nucleons in nuclei and astrophysical systems. Nuclear
interactions are effective interactions for the relevant energies, and 
present one rung in the ladder of effective theories. Nuclear systems 
should therefore not be thought of as a simple extension of hadronic 
physics to lower energies, but rather as a separate many-body problem 
with distinct degrees of freedom. The different effective theories should 
not be thought of as more or less fundamental. What drives the physics 
at one level is frequently not the details of the next level, but rather 
the general aspects of symmetries, renormalization and many-body physics. 
What is important in understanding nuclear physics in the laboratory 
and in stars is not the EFT at some higher energy scale, but rather 
the effective interaction at the energies of the phenomena of interest.

Nuclear interactions have been derived in an EFT for low-energy QCD.
They depend on the cutoff or the resolution scale of the effective 
theory and the renormalization is identical to the running couplings 
in quantum field theories. The RG can be used to change the resolution 
scale in nuclear interactions, and we have shown how all microscopic 
nuclear forces evolve to a universal NN interaction for nucleons with 
momenta $k \lesssim 2.1 \fmi$. In contrast to the different potential
models used traditionally in nuclear physics, we now have a
low-momentum interaction that depends on the resolution scale.
In addition, nuclear EFT offers for the first time a consistent and 
practical expansion scheme for many-body interactions and the coupling
to electromagnetic/weak probes. As a result, calculations with 
microscopic 3N forces beyond the lightest nuclei are now possible.

New frontiers in nuclear physics will be set by model-independent 
predictions of nuclear structure and nucleonic matter. The nuclear 
many-body problem shares many of the approaches and methods with 
atomic, condensed matter and high-energy many-body problems. All 
modern advances in nuclear many-body physics reach over physics 
subfield barriers: for example, EFT for few-nucleon systems and cold 
atoms, the Coupled Cluster method and DFT for nuclear and electronic 
structure, and the RG for superfluidity in neutron stars and 
low-dimensional Fermi systems. While nuclear physics exhibits all 
features of many-body systems, it is distinguished by unique aspects: 
for example, the connection to QCD, its importance to astrophysical systems
and to our very own existence, and its key part to the studies of 
fundamental symmetries. Nuclear physics therefore has and will contribute 
strongly to the general understanding of many-body systems.

\end{document}